\documentclass[11pt,twocolumn]{article}
\usepackage{graphicx}

\begin{document}

\title{\textsf{\textbf{Copernicus's epicycles from Newton's gravitational force law\\ 
via linear perturbation theory in geometric algebra}}}
\author{Quirino M. Sugon Jr.,* Sarah Bragais, and Daniel J. McNamara
\smallskip\\
\small{Ateneo de Manila University, Department of Physics, Loyola Heights, Quezon City, Philippines 1108}\\
\small{*Also at Manila Observatory, Upper Atmosphere Division, Ateneo de Manila University Campus}\\
\small{e-mail: \texttt{qsugon$@$observatory.ph}}}
\date{\small{17 July 2008}}
\maketitle

\section*{}
\small{\textbf{Abstract.} 
We derive Copernicus's epicycles from Newton's gravitational force law by assuming that a planet's orbit is a perturbed circular orbit, with the perturbation defined to be co-rotating with the said orbit.  We substitute this orbit expression into Newton's gravitation law and showed that the perturbation satisfies the linear part of Hill's oscillator equation for lunar motion.  We solve this oscillator equation using an exponential Fourier series and impose the boundary conditions at the aphelion and perihelion to derive  the Copernicus's formulas for the eccentric, deferent, and epicycle.  We show that for small eccetricity, the Copernican orbit expression also leads to Kepler's law of areas for planetary motion.  The formalism we use is the Clifford (geometric) algebra $\mathcal Cl_{2,0}$.

\section{Introduction}
In many introductory physics courses, especially those dealing
with the history and philosophy of science, the Copernican model
is taught conceptually but not mathematically.\cite{Thurston1994p208-209}  And in undergraduate and graduate physics courses, it
is not even mentioned at all. One possible reason is that, unlike in the case of Kepler's ellipse, Copernicus' epicycles
has not been rigorously derived before from first principles,
i.e., from Newton's laws of motion and gravitation. So our aim in
this paper is to present this derivation.  But before we do so,
let us first review the Copernican model.

Copernicus believed that planets orbit around the sun. If the
orbit of a planet is circular with the sun at the center, then
the planet's position in complex form is
\begin{equation}
\label{eq:circular orbit complex} \hat r=
r_0e^{\hat\imath\omega_0t},
\end{equation}
where $r_0$ and $\omega_0$ are the planet's orbital radius and
frequency, respectively.  But because planets sometimes move
closest to the sun (perihelion) and sometimes farthest (aphelion),
Copernicus displaced the center of the planet's circular orbit
a little away from the sun to a new point called the eccentric, so
that the new position $\hat r$ of the planet is
\begin{equation}
\label{eq:eccentric circular orbit complex} \hat r = r_{-1} +
r_0e^{\hat\imath\omega_0t},
\end{equation}
where $r_{-1}$ is the distance of the eccentric from the sun.

Actually, the eccentric hypothesis in Eq.~(\ref{eq:eccentric
circular orbit complex}) is not Copernicus' original idea but was
already known more than a thousand years prior by Ptolemy (though
he assumed that the earth is at rest and not the sun as in
Copernicus).  In fact, if we factor out the exponential
$e^{\hat\imath\omega_0t}$ in Eq.~(\ref{eq:eccentric circular orbit
complex}), we would arrive at Ptolemy's theorem applied by
Copernicus in his heliocentric theory:
\begin{equation}
\label{eq:eccentric-epicycle equivalence} \hat r = r_{-1} +
r_0e^{\hat\imath\omega_0t}=(r_0 +
r_{-1}\,e^{-\hat\imath\omega_0t})e^{\hat\imath\omega_0t}.
\end{equation}
In Ptolemaic terms, $r_{-1}e^{-\hat\imath\omega_0t}$ is called an
epicycle and Eq.~(\ref{eq:eccentric-epicycle equivalence}) is
called the eccentric-epicycle equivalence theorem.  (The actual
theorem is stated geometrically.\cite{Taliaferro1952p86-88,Wallis1952p653-657}) Notice that the theorem essentially states
the equivalence of the description of the planet's position in the
inertial frame (left hand side) and in the rotating frame
(quantity in parenthesis on the right hand side).

Yet Eq.~(\ref{eq:eccentric circular orbit complex}) is still not
consistent with the numerical data. Ptolemy resolved this problem
by assuming that the planet's circular orbit is uniform not with
respect to the orbit's geometric center but on another point
called the equant\cite{Toomer1998p443,Gallavotti1999p10}. Though this construction saves the appearances, Copernicus claimed that the equant goes against the idea of uniform circular motion and for him this is "not sufficiently pleasing to the mind"\cite{Rosen1937}. To remedy this aesthetic difficulty, Copernicus added on top of his original circle in the inertial frame another circle with twice the frequency\cite{Wallis1952p742-743}.  In complex notation, we write 
\begin{equation}
\label{eq:Copernican orbit intro} \hat r=r_{-1} +
r_0e^{\hat\imath\omega_0t}+r_1e^{2\hat\imath\omega_0t}\nonumber\\ 
\end{equation}
where $r_{-1}=-3r_1$.

In terms of the orbit's semimajor axis and eccentricity~$\epsilon$, the Copernican expression in Eq.~(\ref{eq:Copernican orbit intro}) becomes
\begin{equation}
\label{eq:Copernican orbit intro A epsilon}
\hat r=A(\frac{3}{2}\epsilon +e^{\hat\imath\omega_0t}
-\frac{1}{2}\epsilon\, e^{2\hat\imath\omega_0t}),
\end{equation}
as given by Gallavotti\cite{Gallavotti1999p10}.  Notice that Eq.~(\ref{eq:Copernican orbit intro A epsilon}) is different from that derived from Kepler's elliptical orbit for small eccentricity:\cite{Gallavotti1999p6}
\begin{equation}
\label{eq:Copernican orbit from Kepler}
\hat r=A(\epsilon(1-\hat\imath)+ e^{i\omega_0t}+\epsilon(1+\hat\imath)e^{2i\omega_0t}).
\end{equation}
In this paper, our aim is to show that the Copernican expression in Eq.~(\ref{eq:Copernican orbit intro A epsilon}) is a consequence of Newton's gravitational force law.

We shall divide the paper into four sections.  The first section is
Introduction. In the second section, we shall present a brief
tutorial on the Clifford (geometric) algebra $\mathcal Cl_{2,0}$ for the plane\cite{Hestenes1990p48-53,Jancewicz1988p1-17,DoranandLasenby2003p11-15,Lounesto1996p6-13,Calvet2007}, which combines scalars, vectors and imaginary numbers.  We shall show how the exponential Fourier series are related to eccentrics, deferents, and epicycles.\cite{Hanson1960,Saari1990}  In the third section, we shall introduce a pertubation in the planet's position in the frame co-rotating with the planet's unperturbed circular orbit and substitute the result to the vector form of Newton's law of gravitation.  We shall show that the perturbation in complex form satisfies the linear harmonic oscillator equation
\begin{equation}
\label{eq:Hill linear oscillator equation intro} 0=\ddot{\hat
s}+2\hat\imath\omega_0\dot{\hat s}-\frac{3}{2}\omega_0^2(\hat
s+\hat s^*),
\end{equation}
whose scalar and imaginary parts are
\begin{eqnarray}
\label{eq:scalar Hill linear oscillator equation phi is 0}
0&=&\ddot{x}_s-2\omega_0\dot{y}_s-3\omega_0^2x_s,\\
\label{eq:imaginary Hill linear oscillator equation phi is 0}
0&=&\ddot{y_s}+2\omega_0\dot{x}_s.
\end{eqnarray}
These equations are the linear part of Hill's equations for lunar motion\cite{Hill1878p14,Szebehely1967p607}.  We shall solve Eq.~(\ref{eq:Hill linear oscillator equation intro}) using exponential Fourier series and impose the boundary conditions at the aphelion and perihelion to derive the Fourier coefficients of the Copernican orbit.  And the fourth section is Conclusions.

\section{Geometric Algebra}

\subsection{Vectors and Complex Numbers}
The Clifford (geometric) algebra $\mathcal{C}l_{2,0}$ is an
associative algebra generated by two vectors $\mathbf e_1$ and
$\mathbf e_2$ that correspond to the basis vectors along the $x-$
and $y-$axis in the Cartesian coordinate system.  The vectors
satisfy the orthonormality relation
\begin{equation}
\label{eq:orthonormality}
\mathbf e_\mu\mathbf e_\nu+\mathbf e_\nu\mathbf e_\mu=2\delta_{\mu\nu},
\end{equation}
for $\mu,\nu=1,2$.  That is,
\begin{eqnarray}
\label{eq:normality axiom}
\mathbf e_1^2=\mathbf e_2^2=1,\\
\label{eq:orthogonality axiom}
\mathbf e_1\mathbf e_2=\mathbf e_2\mathbf e_1.
\end{eqnarray}
The first equation algebraically defines $\mathbf{e}_1$ and
$\mathbf{e}_2$ as unit vectors by setting their squares to unity; the second equation defines the vectors as mutually orthogonal by making their product anticommute.

Let us define the unit bivector 
\begin{equation}
\label{eq:i}
\hat\imath = \mathbf e_1\mathbf e_2.
\end{equation}
From the orthonormality axiom in
Eq.~(\ref{eq:orthonormality}), it is easy to see that $\hat\imath$
is an imaginary number,
\begin{equation}
\label{eq:i squared}
\hat\imath^2=\mathbf e_1\mathbf e_2\mathbf e_1\mathbf e_2=-\mathbf e_1(\mathbf e_2\mathbf e_2)\mathbf e_1=-\mathbf e_1\mathbf e_1=-1,
\end{equation}
that anticommutes with vectors $\mathbf
e_1$ and $\mathbf e_2$:
\begin{eqnarray}
\label{eq:e1 i anticommutation}
\mathbf e_1\hat\imath &=& \mathbf e_2=-\hat\imath\mathbf e_1,\\
\label{eq:e2 i anticommutation}
\mathbf e_2\hat\imath &=& -\mathbf e_1=-\hat\imath\mathbf e_2.
\end{eqnarray}
Notice that right-multiplying $\hat\imath$ to a vector rotates it
counterclockwise by $\pi/2$.

In general, a vector $\mathbf a$ in the two-dimensional space
spanned by $\mathbf e_1$ and $\mathbf e_2$ is given by
\begin{equation}
\label{eq:vector a complex and conjugate} \mathbf a=a_x\mathbf
e_1+a_y\mathbf e_2 = \mathbf e_1\hat a=\hat a^*\mathbf e_1,
\end{equation}
where
\begin{eqnarray}
\label{eq:a}
\hat a &=&a_x+a_y\hat\imath,\\
\label{eq:a conjugate}
\hat a^* &=&a_x-a_y\hat\imath.
\end{eqnarray}
Equations~(\ref{eq:vector a complex and conjugate}) to (\ref{eq:a conjugate}) relates the vector
$\mathbf a$ to the complex number $\hat a$ and its complex
conjugate $\hat a^*$.

If vector $\mathbf b=b_x\mathbf e_1+b_y\mathbf e_2$, then the
product of vectors $\mathbf a$ and $\mathbf b$ is
\begin{equation}
\label{eq:ab product} \mathbf a\mathbf b=\mathbf a\cdot\mathbf
b+\mathbf a\wedge\mathbf b =\hat a^*\hat b,
\end{equation}
where
\begin{eqnarray}
\label{eq:a dot b}
\mathbf a\cdot\mathbf b &=& a_xb_x+a_yb_y,\\
\label{eq:a wedge b}
\mathbf a\wedge\mathbf b &=& (a_xb_y-a_yb_x)\hat\imath
\end{eqnarray}
are the scalar (dot) and imaginary (bivector or planar) parts of
the product $\mathbf a\mathbf b=\hat a^*\hat b$.  Notice
that the magnitude of the wedge product is that of the cross
product $\mathbf a\times\mathbf b$.  (Geometrically, we say that $\mathbf a\times\mathbf b$ is the vector perpendicular to the oriented plane $\mathbf a\wedge\mathbf b$, though technically, $\mathbf a\times\mathbf b$ is not defined in $\mathcal Cl_{2,0}$---only in $\mathcal Cl_{3,0}$).

\subsection{Circles, Epicycles, and Fourier Series}

Because $\hat\imath$ is an imaginary number, then Euler's theorem
holds:
\begin{equation}
\label{eq:Euler}
e^{\hat\imath\theta}=\cos\theta+\hat\imath\sin\theta,
\end{equation}
where $\theta$ is a real number.  If we left-multiply
Eq.~(\ref{eq:Euler}) by $\mathbf e_1$, we get
\begin{equation}
\label{eq:e1 exp i theta} 
\mathbf e_1e^{\hat\imath\theta}=\mathbf e_1\cos\theta+\mathbf e_2\sin\theta,
\end{equation}
where we used Eq.~(\ref{eq:e1 i anticommutation}).  Equation~(\ref{eq:e1 exp i theta}) states that $\mathbf
e_1e^{\hat\imath\theta}$ is the vector $\mathbf e_1$ rotated
counterclockwise by an angle $\theta$ (assuming that $\mathbf e_1$
points to the right and $\mathbf e_2$ points up).

\begin{figure}[h]
\label{fig:uniform circular motion}
\begin{center}
\setlength{\unitlength}{1 mm}
\begin{picture}(60,60)(-25,-22)
\qbezier(25,0)(24.999,10.356)(17.678,17.678)
\qbezier(17.678,17.678)(10.355,25.000)(0,25)
\qbezier(0,25)(-10.355,25.000)(-17.678,17.678)
\qbezier(-17.678,17.678)(-24.999,10.356)(-25.000,0.000)
\qbezier(-25,0)(-24.999,-10.356)(-17.678,-17.678)
\qbezier(-17.678,-17.678)(-10.355,-25.000)(0,-25)
\qbezier(0,-25)(10.355,-25.000)(17.678,-17.678)
\qbezier(17.678,-17.678)(24.999,-10.356)(25,0)
\qbezier(0,0)(30,0)(30,0)
\qbezier(30,0)(28,-1)(28,-1)
\qbezier(30,0)(28,1)(28,1)
\put(31,-1){$\mathbf e_1$}
\qbezier(0,0)(0,30)(0,30)
\qbezier(0,30)(1,28)(1,28)
\qbezier(0,30)(-1,28)(-1,28)
\put(-1,32){$\mathbf e_2$}
\qbezier(0,0)(24.148,6.470)(24.148,6.470)
\put(10,0.5){$\phi$}
\qbezier(5,0)(5.000,2.887)(2.500,4.330)
\thicklines
\qbezier(0,0)(12.500,21.651)(12.500,21.651)
\thinlines
\qbezier(12.5,19)(12.500,21.651)(12.500,21.651)
\qbezier(10,20)(12.500,21.651)(12.500,21.651)
\put(7,5){$\omega t$}
\put(14,23){$\mathbf r$}
\put(4,12){$r$}
\end{picture}
\end{center}
\end{figure}
\begin{quote}
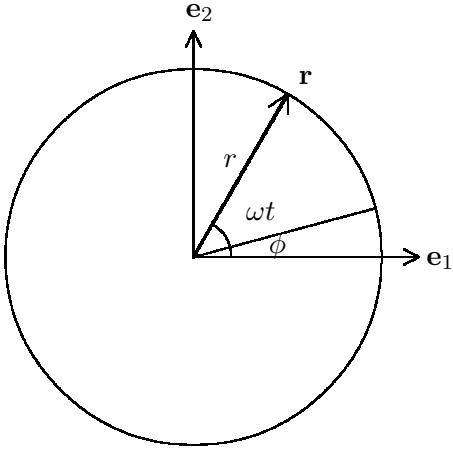
\textbf{Fig.~1}.  The vector $\mathbf r=\mathbf e_1 re^{i(\omega t+\phi)}$.
\end{quote}

The theorem in Eq.~(\ref{eq:e1 exp i theta}) enables us to
express the position $\mathbf r$ of a point in uniform circular
motion as
\begin{equation}
\label{eq:circular motion} \mathbf r=\mathbf
e_1re^{\hat\imath(\omega t+\phi)}=\mathbf e_1r\cos(\omega
t+\phi)+\mathbf e_2r\sin(\omega t+\phi),
\end{equation}
where $r$ is radius, $\omega$ is the angular frequency, and $\phi$
is the rotational phase angle.  Another way to express $\mathbf r$ is
\begin{equation}
\label{eq:r is e1 r psi} 
\mathbf r=\mathbf e_1\hat
r\hat\psi,
\end{equation}
where
\begin{eqnarray}
\label{eq:r is r exp i phi}
\hat r &=& re^{\hat\imath\phi},\\
\label{eq:psi is exp i omega t}
\hat\psi &=& e^{\hat\imath\omega t}
\end{eqnarray}
are the complex radius and rotor (rotation operator),
respectively.  (See Fig.~(1))

Let $\mathbf r_1$ and $\mathbf r_2$ be two rotating vectors:
\begin{eqnarray}
\label{eq:r1 is e1 r1 psi1}
\mathbf r_1&=&\mathbf e_1\hat r_1\hat\psi_1=\mathbf e_1r_1e^{i(\omega_1t+\phi_1)},\\
\label{eq:r2 is e1 r2 psi2}
\mathbf r_2&=&\mathbf e_1\hat r_2\hat\psi_2=\mathbf e_1r_2e^{i(\omega_2t+\phi_2)}.
\end{eqnarray}
If we displace their sum by a vector $\mathbf r_0$,
\begin{equation}
\label{eq:r0 is e1 r0 exp i phi0} \mathbf r_0=\mathbf e_1\hat r_0=\mathbf e_1r_0e^{\hat\imath\phi_0},
\end{equation}
we arrive at
\begin{eqnarray}
\label{eq:r vector is eccentric deferent epicycle} 
\mathbf r=\mathbf e_1(\hat r_0+\hat r_1\hat\psi_1+\hat r_2\hat\psi_2).
\end{eqnarray}

One way to simplify Eq.~(\ref{eq:r vector is eccentric deferent epicycle}) is to set $\omega_1=\omega$ and $\omega_2=2\omega$.  So using the definition of the rotor $\psi$ in Eq.~(\ref{eq:psi is exp i omega t}), we get
\begin{equation}
\label{eq:r vector is Fourier k equals 0 1 2} 
\mathbf r=\mathbf e_1(\hat
r_0+\hat r_1\hat\psi+\hat r_2\psi^2).
\end{equation}
The zeroth harmonic is the eccentric; the first, the deferent; and the second, the epicycle.  In general, we may express the position $\mathbf r$ in time $t$ as
an infinite Fourier series:
\begin{equation}
\label{eq:r is Fourier e1 r psi} 
\mathbf r=\mathbf
e_1\sum_{k\,=-\infty}^\infty \hat
r_k\hat\psi^k=\sum_{k\,=-\infty}^\infty\mathbf e_1re^{\hat\imath(k\omega t+\phi_k)}.
\end{equation}
Equation~(\ref{eq:r is Fourier e1 r psi}) represents the Copernican ideal of decomposing an orbit as a sum of epicycles with harmonic frequencies.

\section{Copernican Dynamics}

\subsection{Uniform Circular Orbit}
In Newton's law of gravitation, the equation of motion of a planet
of mass $m$ revolving around the sun of mass $M$ is
\begin{equation}
\label{eq:Newton gravitation law} \ddot{\mathbf
r}=-GM\frac{\mathbf r}{|\mathbf r|^3},
\end{equation}
where $\mathbf r$ is the position of the planet with respect to the the sun at the origin. 

One to solution to Eq.~(\ref{eq:Newton gravitation law}) is a circular orbit:
\begin{equation}
\label{eq:r vector circular} \mathbf r= \mathbf r_0=\mathbf
e_1\hat r_0\hat\psi_0=\mathbf e_1r_0e^{\hat\imath(\omega_0
t+\phi_0)},
\end{equation}
where $r_0$ is the orbital radius, $\omega_0$ is the orbital frequency, and $\phi_0$ is the orbital phase angle.  

To verify this claim, we first take the derivatives in time of the position vector $\mathbf
r$:
\begin{eqnarray}
\label{eq:r vector circular dot}
\dot{\mathbf r} &=& \omega_0\mathbf e_1\hat\imath\hat
r_0\hat\psi_0=-\hat\imath\omega_0\mathbf r_0,\\
\label{eq:r vector circular dot dot}
\ddot{\mathbf r} &=&-\omega_0^2\mathbf r_0.
\end{eqnarray}
Next, we take the square of the position $\mathbf r$ by using the
conjugation theorem in Eq.~(\ref{eq:vector a complex and
conjugate}):
\begin{equation}
\label{eq:r vector circular squared}
 \mathbf r^2 = \mathbf e_1\hat
r_0\hat\psi_0\mathbf e_1\hat r_0\hat\psi_0 = \hat r_0^*
\hat\psi_0^{-1}\hat r_0\hat\psi_0 = \hat r_0^* \hat
r_0=r_0^2,
\end{equation}
so that $|\mathbf r|=r_0$ as we expect.  And finally, we
substitute Eqs.~(\ref{eq:r vector circular dot}) to
(\ref{eq:r vector circular squared}) back to Eq.~(\ref{eq:Newton
gravitation law}) to arrive at 
\begin{equation}
\label{eq:circular orbit condition} \omega_0^2=\frac{GM}{r_0^3},
\end{equation}
which is the circular orbit condition.

\subsection{Linear Perturbation Theory}

Let us assume that the solution to Eq.~(\ref{eq:Newton gravitation
law}) may be expressed as a sum of a circular orbital position
$\mathbf r_0$ and its small correction $\mathbf r_1$:
\begin{equation}
\label{eq:r is r0 + lambda r1} 
\mathbf r=\mathbf r_0+\lambda\mathbf r_1,
\end{equation}
where $\lambda$ is a perturbation parameter that will be set to
unity later.  If we also assume that the perturbation $\mathbf r_1$
lies in the same orbital plane as the original circular orbit
$\mathbf r_0$ in Eq.~(\ref{eq:r vector circular}) and
co-rotating with it, then we may write $\mathbf r_1$ as
\begin{equation}
\label{eq:r1 is e1 s psi0} 
\mathbf r_1=\mathbf e_1\hat s\hat\psi_0,
\end{equation}
where $\hat s$ is a complex function.  Hence,
\begin{equation}
\label{eq:r vector perturbed} 
\mathbf r=\mathbf e_1(\hat r_0+\lambda\hat s)\hat\psi_0.
\end{equation}

Taking the first and second time derivatives of the
position $\mathbf r$ in Eq.~(\ref{eq:r vector perturbed}),
we get
\begin{eqnarray}
\label{eq:r vector perturbed dot}
\dot{\mathbf r}&=&\mathbf e_1(\lambda\dot{\hat
s}+\hat\imath\omega_0(\hat r_0+\lambda\hat s))\hat\psi_0,\\
\label{eq:r vector perturbed dot dot}
\ddot{\mathbf r}&=&\mathbf e_1(\lambda\ddot{\hat
s}+2\lambda\hat\imath\omega_0\dot{\hat s}-\omega_0^2(\hat
r_0+\lambda\hat s))\hat\psi_0.
\end{eqnarray}
Equation~(\ref{eq:r vector perturbed dot dot}) provides the expansion of the left side of Newton's gravitation law in Eq.~(\ref{eq:Newton gravitation law}).

On the other hand, to rewrite the right side of the gravitation
law, we need first to take the square of the position vector
$\mathbf r$ in Eq.~(\ref{eq:r vector perturbed}) and
retain only the terms up to first order in $\lambda$:
\begin{equation}
\label{eq:r vector perturbed squared} 
\mathbf r^2=(\hat r_0+\lambda\hat s)^*(\hat r_0+\lambda\hat s)\approx r_0^2+\lambda(\hat r_0^*\hat s + \hat r_0\hat s^*).
\end{equation}
Raising both sides of Eq.~(\ref{eq:r vector perturbed squared}) to $-3/2$ power and employing the binomial theorem, we get
\begin{equation}
\label{eq:r vector magnitude inverse cube} \frac{1}{|\mathbf
r|^3}\approx
\frac{1}{r_0^3}\left(1-\lambda\frac{3}{2r_0}(\hat\eta_0^*\hat
s+\hat\eta_0\hat s^*)\right),
\end{equation}
where
\begin{equation}
\label{eq:eta0 is exp i phi0} \hat\eta_0=e^{\hat\imath\phi_0}.
\end{equation}
Multiplying Eq.~(\ref{eq:r vector magnitude inverse cube}) by the
position $\mathbf r$ in Eq.~(\ref{eq:r vector perturbed}) yields
\begin{equation}
\label{eq:perturbed position over magnitude cube} \frac{\mathbf
r}{|\mathbf r|^3}\approx \frac{1}{r_0^3}\mathbf r_0+\mathbf
e_1\frac{\lambda}{r_0^3}\left(\hat s-\frac{3}{2r_0}(\hat s
+\hat\eta_0^2\hat s^*)\right)\hat\psi_0,
\end{equation}
where we retained only the terms up to first order in $\lambda$. 

Now, substituting Eqs.~(\ref{eq:r vector perturbed})
and ~(\ref{eq:r vector magnitude inverse cube}) back to the
gravitation law in Eq.~(\ref{eq:Newton gravitation law}), we
arrive at
\begin{equation}
\label{eq:Hill linear oscillator equation}
0=\ddot{\hat
s}+2\hat\imath\omega_0\dot{\hat s}-\frac{3}{2}\omega_0^2(\hat
s+\hat\eta_0^2\hat s^*).
\end{equation}
If we set $\phi_0=0$ (this means that orbit is not tilted, as we shall show later), so that $\hat\eta_0=e^{i\phi_0}=1$, we get Hill's oscillator equation in Eq.~(\ref{eq:Hill linear oscillator equation intro}).

\section{Copernican Analysis}

\subsection{Epicyclical Fourier Series}

We assume that the solution to the orbital harmonic oscillator
equation in Eq.~(\ref{eq:Hill linear oscillator equation}) is an
exponential Fourier series with $\omega_0$ as the fundamental angular frequency:
\begin{equation}
\label{eq:s is Fourier a psi0} 
\hat s=\sum_{k\,=-\infty}^\infty\hat a_k\hat\psi_0^k.
\end{equation}
The time derivatives of $\hat s$ are
\begin{eqnarray}
\label{eq:s is Fourier a psi0 dot}
\dot{\hat s}&=&\hat\imath\omega_0\sum_{k\,=-\infty}^\infty k\hat
a_k\hat\psi_0^k,\\
\label{eq:s is Fourier a psi0 dot dot}
\ddot{\hat s} &=& -\omega_0^2\sum_{k\,=-\infty}^\infty k^2\hat
a_k\hat\psi_0^k,
\end{eqnarray}
while the conjugate of $\hat s$ is
\begin{equation}
\label{eq:s is Fourier a psi0 conjugate} 
\hat s^*=\sum_{k\,=-\infty}^\infty\hat
a_k^*\hat\psi_0^{-k}=\sum_{k\,=-\infty}^\infty \hat
a_{-k}^*\hat\psi_0^k.
\end{equation}

Substituting Eqs.~(\ref{eq:s is Fourier a psi0}) to
~(\ref{eq:s is Fourier a psi0 conjugate}) back to Eq.~(\ref{eq:Hill linear oscillator equation}), we get
\begin{equation}
\label{eq:s Fourier substituted to Hill linear oscillator equation}
0=\sum_{k\,=-\infty}^\infty((k^2+2k+\frac{3}{2})\hat
a_k+\frac{3}{2}\hat\eta_0^2\hat a_k^*)\hat\psi_0^k,
\end{equation}
after factoring out $-\omega_0^2$ and rearranging the terms.
Because the rotors $\hat\psi_0^k$ are orthonormal in the Fourier
sense, then Eq.~(\ref{eq:s Fourier substituted to Hill linear oscillator equation}) holds  only if the coefficient of $\hat\psi_0^k$ is zero for all $k$:
\begin{equation}
\label{eq:Fourier coefficient relation} 
0=(k^2+2k+\frac{3}{2})\hat a_k +\frac{3}{2}\hat\eta_0^2\hat a_k^*.
\end{equation}

Solving for the coefficient $\hat a_k$ in Eq.~(\ref{eq:Fourier
coefficient relation}), we get
\begin{equation}
\label{eq:a_-k is function of a_k} 
\hat a_{-k}=-\frac{2}{3}\hat\eta_0^2(k^2+2k+\frac{3}{2})\hat
a_k^*.
\end{equation}
Replacing the index $k$ by $-k$,
\begin{equation}
\label{eq:a_k is function of a_-k}
a_k=-\frac{2}{3}\hat\eta_0^2(k^2-2k+\frac{3}{2})\hat
a_{-k}^*,
\end{equation}
and substituting the result back in Eq.~(\ref{eq:a_-k is function of a_k}), we arrive at
\begin{equation}
\label{eq:k quadratic polynomial constraint}
0=(k^2+2k+\frac{3}{2})(k^2-2k+\frac{3}{2})-\frac{9}{4}=k^2(k^2-1),
\end{equation}
after factoring out $\hat a_k$ and rearranging the terms.  Hence,
\begin{equation}
\label{eq:k is -1 0 1} 
k=\{-1,0,1\}.
\end{equation}

Because the values of the index $k$ are limited by
Eq.~(\ref{eq:k is -1 0 1}), then the Fourier series for
the perturbation $\hat s$ in Eq.~(\ref{eq:s is Fourier a psi0}) simplifies to
\begin{equation}
\label{eq:s with k equals -1 0 1} 
\hat s=\hat a_{-1}\hat\psi_0^{-1}+\hat a_0+\hat a_1\hat\psi_0.
\end{equation}
The relationship between the coefficients $\hat a_{-1}$ and $\hat
a_1$ in Eq.~(\ref{eq:s with k equals -1 0 1}) may be
obtained by setting $k=1$ in Eq.~(\ref{eq:a_-k is function of a_k}):
\begin{equation}
\label{eq:a_-1 is function of a_1} 
\hat a_{-1}=-3\hat\eta_0^2\hat a_1^*.
\end{equation}
Similarly, the condition for $\hat a_0$ is
\begin{equation}
\label{eq:a_0} 
\hat a_0=-\hat\eta_0^2\hat a_0^*.
\end{equation}
This is satisfied in three possible ways:
\begin{equation}
\label{eq:a_0 as i eta_0 or 0} \hat a_0 =
\{\pm\hat\imath\hat\eta_0,0\}.
\end{equation}

Substituting the expression for $\hat s$ in Eq.~(\ref{eq:s with k equals -1 0 1}) back to the position vector expression in Eq.~(\ref{eq:r vector perturbed}), we get
\begin{equation}
\label{eq:r vector with k equals -1 0 1} 
\mathbf r=\mathbf e_1(\hat a_{-1}+(\hat r_0+\hat a_0)\hat\psi_0+\hat a_1\hat\psi_0^2).
\end{equation}
Let us count the number of unknowns in this equation.  The
coefficient $\hat a_{-1}$ is related to $\hat a_1$ by
Eq.~(\ref{eq:a_-1 is function of a_1}).  The angular frequency
$\omega_0$ in $\hat\psi_0=e^{i\omega_0 t}$ is related to the radius $r_0$ of $\hat r_0=r_0\hat\eta_0=r_0e^{i\phi_0}$ by Eq.~(\ref{eq:circular orbit condition}).  The phase angle $\phi_0$ of $\hat r_0$ is related to that of $\hat a_0$ by Eq.~(\ref{eq:a_0 as i eta_0 or 0}).  Thus, there are five unknowns in Eq.~(\ref{eq:r vector with k equals -1 0 1}): $a_{1x},a_{1y},r_0,\phi_0$, and $a_0$.

However, the orbit of a planet in the plane is completely
specified in two ways: (a) given the position $\mathbf r_1$ and
the velocity $\mathbf v_1$ at a particular time $t_1$ or (b) given
the positions $\mathbf r_1$ and $\mathbf r_2$ at their respective
times $t_1$ and $t_2$.  In other words, there are two constraint
vector equations that are equivalent to four scalar equations for
the components. These four equations can only determine four
unknowns and not five, so one of our unknowns is superfluous and
this must be $a_0$ because $\hat a_0=0$ is a possibility in
Eq.~(\ref{eq:a_0 as i eta_0 or 0}).  Thus, Eq.~(\ref{eq:r vector with k equals -1 0 1}) reduces to
\begin{equation}
\label{eq:r vector with k equals -1 0 1 reduced} 
\mathbf r=\mathbf e_1(\hat a_{-1}+\hat r_0\hat\psi_0+\hat
a_1\hat\psi_0^2).
\end{equation}
Because of the similarity of Eq.~(\ref{eq:r vector with k equals -1 0 1 reduced}) to Eq.~(\ref{eq:r vector is Fourier k equals 0 1 2}),
we recognize $\mathbf e_1\hat a_{-1}$ as the eccentric, $\mathbf
e_1\hat r_0\hat\psi_0$ as the deferent, and $\mathbf e_1\hat
a_1\hat\psi_0^2$ as the epicycle in the Copernican model.

\subsection{Boundary Conditions: Aphelion and Perihelion}

Suppose that at $t=0$, the planet is at its aphelion position 
$\mathbf r_a$ at a distance $r_a$ from the sun at a
counterclockwise angle $\gamma$ from the positive $x-$axis; while
at $t=\tau/2=\pi/\omega_0$ the planet is at its perihelion
position $\mathbf r_p$ at a distance $r_p$ from the sun at a
similar angle from the negative $x-$axis.  Imposing these boundary
conditions on the position vector $\mathbf r$ in Eq.~(\ref{eq:r vector with k equals -1 0 1 reduced}) yields two simultaneous
equations:
\begin{eqnarray}
\label{eq:r_a is aphelion}
\mathbf r_a &=& \mathbf e_1 r_ae^{\hat\imath\gamma}=\mathbf
e_1(\hat a_{-1}+\hat r_0+\hat a_1),\\
\label{eq:r_p is perihelion}
\mathbf r_p &=& -\mathbf e_1 r_pe^{\hat\imath\gamma}=\mathbf
e_1(\hat a_{-1}-\hat r_0+\hat a_1).
\end{eqnarray}
Factoring out $\mathbf e_1$ from Eqs.~(\ref{eq:r_a is aphelion}) and (\ref{eq:r_p is perihelion}) and using the expression for $\hat a_{-1}$ in Eq.~(\ref{eq:a_-1 is function of a_1}), we get
\begin{eqnarray}
\label{eq:r_a exp i gamma}
r_ae^{\hat\imath\gamma}&=&-3\hat\eta_0^2\hat a_1^*+\hat r_0+\hat a_1,\\
\label{eq:r_p exp i gamma}
-r_pe^{\hat\imath\gamma}&=&-3\hat\eta_0^2\hat a_1^*-\hat
r_0+\hat a_1,
\end{eqnarray}
which are two simultaneous equations for $\hat r_0$ and $\hat
a_1$.

\textbf{Deferent.}  To solve for $\hat r_0$, we take the difference of Eqs.~(\ref{eq:r_a exp i gamma}) and (\ref{eq:r_p exp i gamma}) to get
\begin{equation}
\label{eq:deferent r_0 is half r_a + r_p exp i gamma} 
\hat r_0 = r_0\hat\eta_0= r_0e^{\hat\imath\phi_0}= \frac{1}{2}(r_a+r_p)e^{\hat\imath\gamma},
\end{equation}
so that
\begin{eqnarray}
\label{eq:r_0 is half r_a + r_p}
r_0&=&\frac{1}{2}(r_a+r_p),\\
\label{eq:phi_0 is gamma}
\phi_0&=&\gamma.
\end{eqnarray}
Thus, the radius $r_0$ of the deferent circle is the length of the semimajor axis of the orbit; the phase angle $\phi_0$ is the angle of inclination of the semimajor axis from the $x-$axis along $\mathbf e_1$.

\textbf{Epicycle.}  To solve for the coefficient $\hat a_1$, we add the Eqs.~(\ref{eq:r_a is aphelion}) and (\ref{eq:r_p is perihelion}) to obtain
\begin{equation}
\label{eq:half r_a - r_p exp i gamma}
\frac{1}{2}(r_a-r_p)e^{\hat\imath\gamma}=-3\hat\eta_0^2\hat
a_1^*+\hat a_1.
\end{equation}
Because $\hat a_1=a_x+\hat\imath a_y$ cannot be readily isolated, we separate the real and imaginary parts of Eq.~(\ref{eq:half r_a - r_p exp i gamma}) to get
\begin{eqnarray}
\label{eq:half r_a - r_p cos gamma}
\frac{1}{2}(r_a-r_p)\cos\gamma &=& a_{1x}(-3\cos 2\gamma+1)\nonumber\\ 
& &+\ a_{1y}(-3\sin 2\gamma),\\
\label{eq:half r_a - r_p sin gamma}
\frac{1}{2}(r_a-r_p)\sin\gamma &=& a_{1x}(-3\sin
2\gamma)\nonumber\\
& &+\ a_{1y}(3\cos 2\gamma + 1),
\end{eqnarray}
where we used the relation $\phi_0=\gamma$ in Eq.~(\ref{eq:phi_0 is gamma}).  Solving for the components
$a_{1x}$ and $a_{1y}$ and combining the results, we arrive at
\begin{equation}
\label{eq:epicycle a_1 is fourth r_a - r_p exp i gamma} 
\hat a_1=-\frac{1}{4}(r_a-r_p)e^{\hat\imath\gamma}.
\end{equation}
Equation~(\ref{eq:epicycle a_1 is fourth r_a - r_p exp i gamma}) states that the radius $a_1$ of the epicycle $\hat a_1$ is one-fourth the difference between the aphelion distance $r_a$ and the perihelion distance $r_p$.  Note the negative sign.

\textbf{Eccentric.}  After knowing $\hat a_1$, we use the coefficient relation in
Eq.~(\ref{eq:a_-1 is function of a_1}) to solve for $\hat a_{-1}$:
\begin{equation}
\label{eq:eccentric a_-1 is three fourths r_a - r_p exp i gamma} \hat a_{-1}=\frac{3}{4}(r_a-r_p)e^{\hat\imath\gamma}.
\end{equation}
Equation~(\ref{eq:eccentric a_-1 is three fourths r_a - r_p exp i gamma}) states that the length $a_{-1}$ of
the eccentric is three-fourth the difference between the aphelion
distance $r_a$ and the perihelion distance $r_p$.

\subsection{Copernican Orbit}

We now substitute the expressions $\hat a-$coefficients in
Eqs.~(\ref{eq:epicycle a_1 is fourth r_a - r_p exp i gamma}) and (\ref{eq:eccentric a_-1 is three fourths r_a - r_p exp i gamma})
and that of $\hat r_0$ in Eq.~(\ref{eq:deferent r_0 is half r_a + r_p exp i gamma}) back to the expression for the position $\mathbf r$ in Eq.~(\ref{eq:r vector with k equals -1 0 1 reduced}) to get
\begin{equation}
\label{eq:r vector is Copernican orbit tilted} 
\mathbf r=\mathbf e_1e^{\hat\imath\gamma} (\frac{3}{4}(r_a-r_p)+\frac{1}{2}(r_a+r_p)\hat\psi_0-\frac{1}{4}(r_a-r_p)\hat\psi_0^2).
\end{equation}
If the semimajor axis' inclination angle $\gamma=\phi_0=0$, then
Eq.~(\ref{eq:r vector is Copernican orbit tilted}) reduces to
\begin{equation}
\label{eq:r vector is Copernican orbit} 
\mathbf r= \mathbf e_1(\frac{3}{4}(r_a-r_p)+\frac{1}{2}(r_a+r_p) \hat\psi_0-\frac{1}{4}(r_a-r_p)\hat\psi_0^2).
\end{equation}
Equation~(\ref{eq:r vector is Copernican orbit}) is our desired approximation of a planet's orbit around the sun using eccentric, deferent, and epicycle in terms of the planet's aphelion $r_a$ and perihelion $r_p$.  (See Fig.~(2))

If we employ the definitions of the Keplerian semimajor axis $A$ and eccentricity $\epsilon$,
\begin{eqnarray}
\label{eq:semimajor axis A}
A &=& \frac{1}{2}(r_a+r_p),\\
\label{eq:eccentricity epsilon}
\epsilon &=& \frac{r_a-r_p}{r_a+r_p}
\end{eqnarray}
then we may rewrite Eq.~(\ref{eq:r vector is Copernican orbit}) as
\begin{equation}
\label{eq:r vector is Copernican orbit A epsilon} 
\mathbf r=\mathbf e_1A(\frac{3}{2}\epsilon +\hat\psi_0
-\frac{1}{2}\epsilon\hat\psi_0^2),
\end{equation}
which is Eq.~(\ref{eq:Copernican orbit intro A epsilon}).  Or in Cartesian coordinates,
\begin{eqnarray}
\label{eq:x is Copernican orbit A epsilon}
x&=&A(\frac{3}{2}\epsilon+\cos(\omega_0t)-\frac{1}{2}\epsilon\cos(2\omega_0t)),\\
\label{eq:y is Copernican orbit A epsilon}
y&=&A(\sin(\omega_0t)-\frac{1}{2}\epsilon\sin(2\omega_0t)).
\end{eqnarray}

\begin{figure}[h]
\label{fig:copernican eccentric deferent epicycle}
\begin{center}
\setlength{\unitlength}{1 mm}
\begin{picture}(60,50)(-20,-25)
\thicklines
\qbezier(32,0)(32,7.239)(28.971,12.971)
\qbezier(28.971,12.971)(24.668,21.111)(16.000,24.000)
\qbezier(16.000,24.000)(4.921,27.693)(-4.971,20.971)
\qbezier(-4.971,20.971)(-16.000,13.475)(-16.000,0.000)

\qbezier(-16.000,0.000)(-16.000,-13.475)(-4.971,-20.971)
\qbezier(-4.971,-20.971)(4.921,-27.693)(16.000,-24.000)
\qbezier(16.000,-24.000)(24.668,-21.111)(28.971,-12.971)
\qbezier(28.971,-12.971)(32,-7.239)(32,0)
\thinlines
\qbezier(0,2)(0,-2)(0,-2)
\qbezier(-2,0)(12,0)(12,0)
\put(4,-4){$a_{-1}$}
\qbezier(12,0)(28.971,16.971)(28.971,16.971)
\put(17,10){$r_0$}
\qbezier(36.000,0.000)(36.000,9.941)(28.971,16.971)
\qbezier(28.971,16.971)(21.941,24.000)(12.000,24.000)
\qbezier(12.000,24.000)(2.059,24.000)(-4.971,16.971)
\qbezier(-4.971,16.971)(-12.000,9.941)(-12.000,0.000)

\qbezier(-12.000,0.000)(-12.000,-9.941)(-4.971,-16.971)
\qbezier(-4.971,-16.971)(2.059,-24.000)(12.000,-24.000)
\qbezier(12.000,-24.000)(21.941,-24.000)(28.971,-16.971)
\qbezier(28.971,-16.971)(36.000,-9.941)(36.000,0.000)
\qbezier(32.971,16.971)(32.971,19.280)(30.971,20.435)
\qbezier(30.971,20.435)(28.971,21.589)(26.971,20.435)
\qbezier(26.971,20.435)(24.971,19.280)(24.971,16.971)

\qbezier(24.971,16.971)(24.971,14.661)(26.971,13.506)
\qbezier(26.971,13.506)(28.971,12.352)(30.971,13.506)
\qbezier(30.971,13.506)(32.971,14.661)(32.971,16.971)
\qbezier(28.971,16.971)(28.971,12.971)(28.971,12.971)
\put(26,10){$a_1$}
\end{picture}
\end{center}
\end{figure}
\begin{quote}
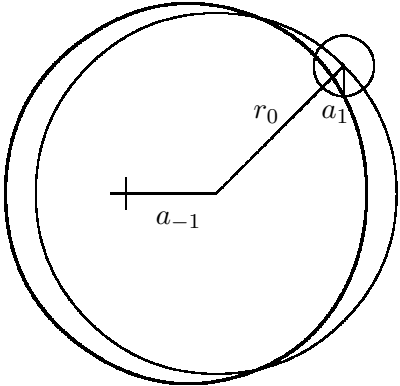
\textbf{Fig.~2}.  Compass-and-straightedge plotting of a Copernican orbit with eccentric distance $a_1$, deferent radius $r_0$, and epicycle radius $a_1$.  The orbit's eccentricity is $\epsilon=1/3$.
\end{quote}


The time derivative of the planet's position $\mathbf r$ in Eq.~(\ref{eq:r vector is Copernican orbit A epsilon}) is
\begin{equation}
\label{eq:v vector is Copernican orbit A epsilon} 
\mathbf v=\mathbf e_1\hat\imath\omega A(\hat\psi_0
-\epsilon\hat\psi_0^2).
\end{equation}
Left-multiplying this by $\mathbf r$,
\begin{eqnarray}
\label{eq:rv Copernican A epsilon}
\mathbf r\mathbf v&=&\hat\imath\omega_0 A^2(\frac{3}{2}\epsilon+\hat\psi_0^{-1}-\frac{1}{2}\epsilon\hat\psi_0^{-2})(\hat\psi_0-\epsilon\hat\psi_0^2)\nonumber\\
&=&\hat\imath\omega_0 A^2[\,-\frac{1}{2}\epsilon\hat\psi_0^{-1} +(1+\frac{1}{2}\epsilon^2)\nonumber\\
& &\qquad\quad+\frac{1}{2}\epsilon\hat\psi_0 -\frac{3}{2}\epsilon^2\hat\psi_0^2\,],
\end{eqnarray}
and separating the scalar and imaginary parts of the result, we arrive at

\begin{eqnarray}
\label{eq:r dot v Copernican A epsilon}
\mathbf r\cdot\mathbf v&=&\omega_0A^2(-\epsilon\sin(\omega_0t)+\frac{3}{2}\epsilon^2\sin(2\omega_0t)),\\
\label{eq:r wedge v Copernican A epsilon}
\mathbf r\wedge\mathbf v&=&\hat\imath\omega_0A^2((1+\frac{1}{2}\epsilon^2)-\frac{3}{2}\epsilon^2\cos(2\omega_0t)).
\end{eqnarray}

\begin{figure}[h]
\label{fig:equal areas in equal times}
\begin{center}
\setlength{\unitlength}{1 mm}
\begin{picture}(60,55)(-22,-22)
\thicklines
\qbezier(32,0)(32,7.239)(28.971,12.971)
\qbezier(28.971,12.971)(24.668,21.111)(16.000,24.000)
\qbezier(16.000,24.000)(4.921,27.693)(-4.971,20.971)
\qbezier(-4.971,20.971)(-16.000,13.475)(-16.000,0.000)

\qbezier(-16.000,0.000)(-16.000,-13.475)(-4.971,-20.971)
\qbezier(-4.971,-20.971)(4.921,-27.693)(16.000,-24.000)
\qbezier(16.000,-24.000)(24.668,-21.111)(28.971,-12.971)
\qbezier(28.971,-12.971)(32,-7.239)(32,0)
\thinlines
\qbezier(0,2)(0,-2)(0,-2)
\put(16,-4){$r_a$}
\qbezier(0,0)(0,0)(32,0)
\qbezier(0,0)(0,0)(30.785,8.536)
\qbezier(0,0)(0,0)(26,17.321)
\qbezier(0,0)(0,0)(30.785,8.536)
\qbezier(0,0)(0,0)(16,24)
\qbezier(0,0)(0,0)(2,24.249)
\qbezier(0,0)(0,0)(-10.785,15.464)
\qbezier(0,0)(0,0)(-16,0)
\put(-8,-4){$r_p$}
\end{picture}
\end{center}
\end{figure}
\begin{quote}
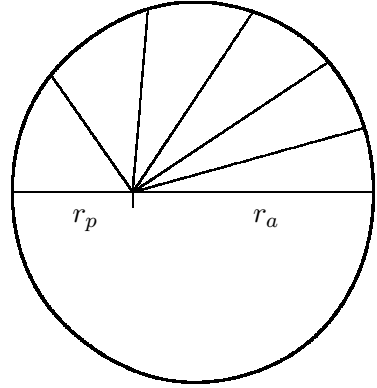
\textbf{Fig.~3}.  The position of a planet in Copernican orbit from $t=0$ to $t=\tau/2$ at a time interval of $\tau/12$.  The orbit's eccentricity is $\epsilon=1/3$.
\end{quote}

For nearly circular orbits, the eccentricity $\epsilon\approx 0$ ($\epsilon$ is $0.0167$ for earth and $0.0068$ for Venus).  So dropping the $\epsilon^2$ terms in Eqs.~(\ref{eq:r dot v Copernican A epsilon}) and (\ref{eq:r wedge v Copernican A epsilon}), we arrive at
\begin{eqnarray}
\label{eq:r dot v Copernican A epsilon linear}
\mathbf r\cdot\mathbf v&=&-\omega_0A^2\epsilon\sin(\omega_0t),\\
\label{eq:r wedge v Copernican A epsilon linear}
\mathbf r\wedge\mathbf v&=&\hat\imath\omega_0A^2.
\end{eqnarray}
The first equation means that the position and velocity of a planet are perpendicular at the aphelion $(t=0)$ and perihelion $(t=\tau/2=\pi/\omega_0)$; the second equation implies that the oriented area
\begin{equation}
\label{eq:A is r wedge v delta t}
A=\mathbf r\wedge(\mathbf v\delta t)=\hat\imath\omega_0A^2\delta t
\end{equation}
swept by the radius vector $\mathbf r$ for a small interval of time $\delta t$ is constant, which is Kepler's second law (we can always perform an integral to show the validity of the law for large time intervals).  (See Fig.~(3))

\section{Summary and Conclusions}
In this paper, we derived the Copernican system of epicycles from Newton's gravitational force law in vector form via linear perturbation theory in Clifford (geometric) algebra $\mathcal Cl_{2,0}$ of the plane.  We assumed that the planet's orbit is a perturbed circular orbit, where the perturbation is defined as a vector co-rotating with the original orbit.  We substituted this expression into Newton's gravitation law.  Using binomial expansion, we showed that this perturbation may be represented by a complex function $\hat s$ that satisfies the linearized form of Hill's equation for lunar motion.  This equation is a linear harmonic oscillator with imaginary damping term and an extra forcing term that is proportional to the conjugate $\hat s^*$.  

We solved this oscillator equation using exponential Fourier series with the frequency $\omega_0$ of the unperturbed circular orbit as the fundamental frequency.  We showed that only three harmonics are allowed: -1, 0, and 1.  This result makes the planet's position as an exponential Fourier series with three harmonics: 0, 1, 2---corresponding to the planet's eccentric, deferent, and epicycle.  We determined the values of the Fourier coefficients by imposing that the planet is at its aphelion at $t=0$ and at its perihelion at $t=\tau/2=\pi/\omega_0$.  And from this we derived Gallavotti's expression for the Copernican orbit in terms of its semimajor axis $A$ and eccentricity $\epsilon$.

We also computed the dot and wedge products of the planet's position and velocity.  We showed that for small eccentricity $\epsilon$, the dot product is proportional to $-\sin(\omega_0t)$; the wedge product is constant, $\hat\imath\omega_0A^2$, which implies that the planet's position vector sweeps out equal areas in equal times, as given by Kepler's second law.

\section*{\small{Acknowledgments}}
This research was supported by the Manila Observatory and by the Physics Department of Ateneo de Manila University.

\end{document}